\documentclass[12pt,a4paper]{article}
\usepackage{graphicx}
\usepackage{times}
\title{\bf Spectroscopic madness---A golden age for amateurs}
\author{Thomas Eversberg\\
\\
\normalsize Schn\"orringen Telescope Science Institute (STScI), Am Kielshof 21a, 51105 K\"oln, Germany
\\
\\
\normalsize Published in proceedings of \\
\normalsize"Stellar Winds in Interaction", editors T. Eversberg and J.H. Knapen. \\ 
\normalsize Full proceedings volume is available on http://www.stsci.de/pdf/arrabida.pdf
}
\date{\mbox{}}
\begin{document}
\maketitle
%
%
\def\bull{\vrule height .9ex width .8ex depth -.1ex}
\makeatletter
\def\ps@plain{\let\@mkboth\gobbletwo
\def\@oddhead{}\def\@oddfoot{\hfil\tiny\bull\quad
Workshop ``Stellar Winds in Interaction'' Convento da Arr\'abida, 2010 May 29 - June 2\quad\bull}%
\def\@evenhead{}\let\@evenfoot\@oddfoot}
\makeatother
%
%
\def\beginrefer{\section*{References}%
\begin{quotation}\mbox{}\par}
\def\refer#1\par{{\setlength{\parindent}{-\leftmargin}\indent#1\par}}
\def\endrefer{\end{quotation}}
%
%
{\noindent\small{\bf Abstract:} 
Today, professional instrumentation is dominated by heavily oversubscribed telescopes which focus mainly on a limited number of ``fashionable" research topics. As a result, time acquisition for massive star research, including extended observation campaigns, becomes more difficult. On the other hand, massive star investigations by amateur astronomers performing spectroscopic measurements are on a level which can fulfil professional needs. I describe the instrumentation available to the amateur, their observational skills and the potential contribution they can make to the professional community.
}
%
%
\section{Introduction}

\begin{figure}[ht]
\centering
\includegraphics[height=6cm]{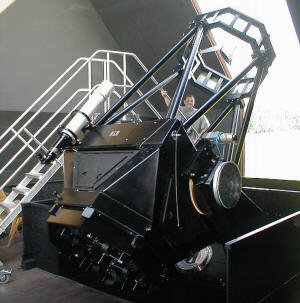}
\caption{\label{one}
The 1.2m Melle telescope is the largest amateur instrument in Germany.}
\end{figure}

\begin{figure}[ht]
\centering
\includegraphics[height=5cm]{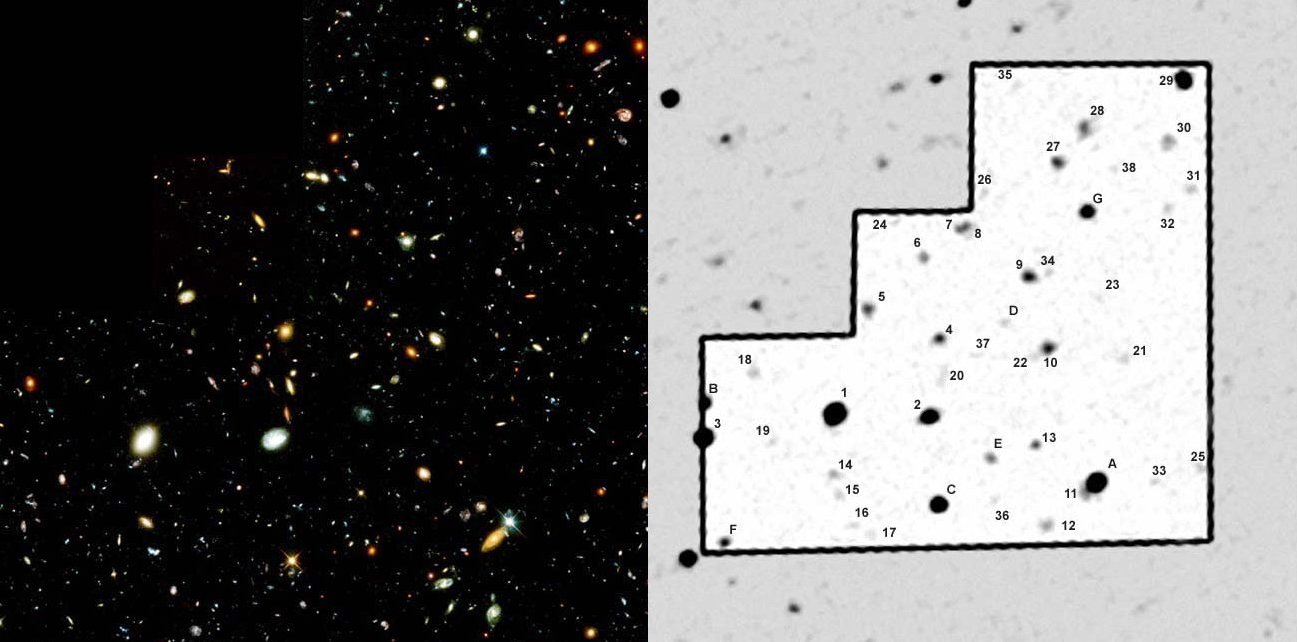}
\caption{\label{deepfield}
The {\it Hubble} Deep Field. Left: {\it HST}. Right: C14, 18\,h exposure time, 10\,km outside Cologne. The faintest objects have about 23 magnitudes in $V$. Courtesy J\"org Zborowska.}
\end{figure}

The so-called ``Golden Age of Astronomy" not only influences professional scientific work but also
the amateur domain. Large instrumentation such as 1m class telescopes have reached the amateur
domain (Fig.~\ref{one}) and optics and CCD detectors are available off-the-shelf for relatively low prices. Today amateurs
can accomplish extraordinary spectroscopic results, which would have been impossible a few decades
ago. For instance, a typical amateur deep sky observer with a small home telescope
can achieve results comparable to modern professional instrumentation, as shown in
Fig.~\ref{deepfield} for the {\it Hubble} Deep Field. Only about two decades ago amateur astronomers began with serious spectroscopic measurements of relatively bright objects, guided by professional needs. Today a worldwide network of spectroscopists with hot spots in France and Germany has been established. It includes instrumental experts who build their own equipment, scientists by education and pure beginners. Information exchange is done via internet (mailing lists, discussion forums, websites) and regular international conferences. It culminates in various professional-amateur (ProAm) projects, mainly on hot stars.    

\section{State-of-the-art amateur spectroscopy}

\begin{figure}[ht]
\centering
\includegraphics[height=6cm]{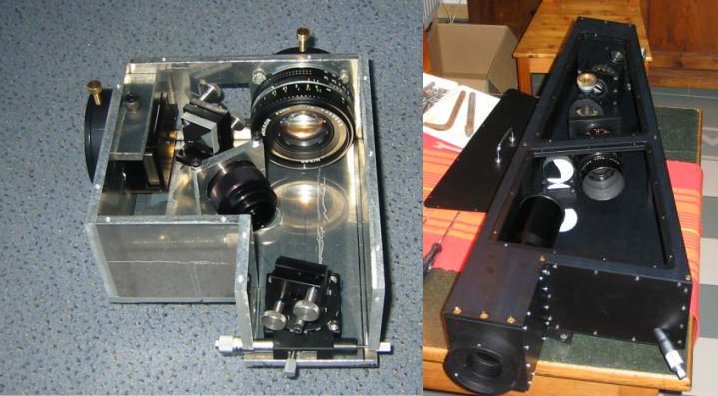}
\caption{\label{specs}
Typical amateur self-made spectrographs. {\it Left}: Device by Bernd Marquardt (Germany). {\it Right}: The maximum efficiency device by the author and Klaus Vollmann (Germany).}
\end{figure}

\begin{figure}[ht]
\centering
\includegraphics[height=5cm]{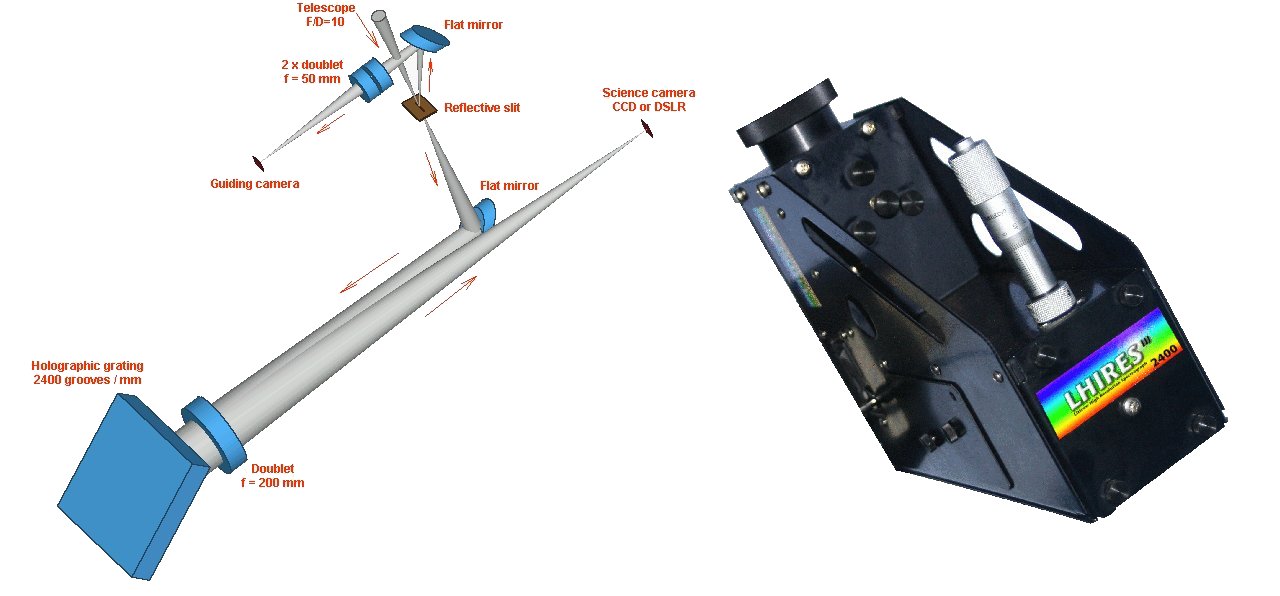}
\caption{\label{lhires}
Off-the-shelf standard Littrow spectrograph LHIRES III from Shelyak.}
\end{figure}

\begin{figure}[ht]
\centering
\includegraphics[height=5cm]{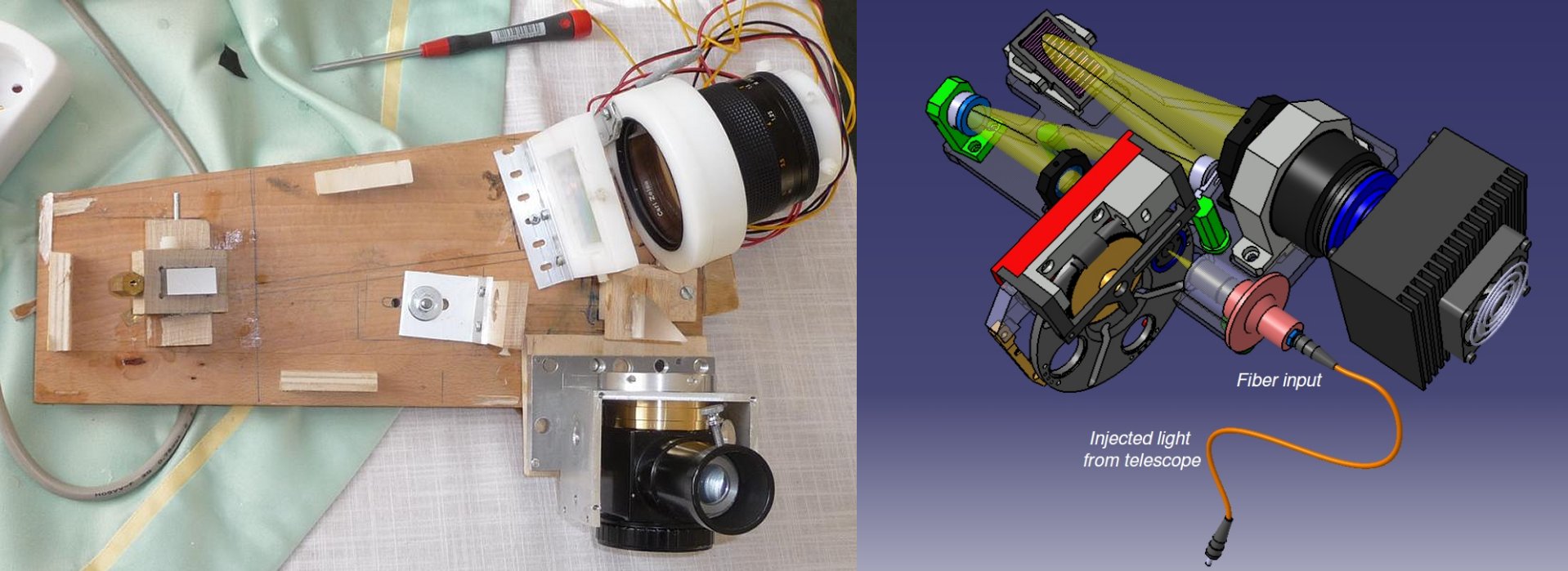}
\caption{\label{stoberfeger}
Two Echelle prototypes both with grating cross-dispersers designed by the amateurs Berthold Stober ({\it left}) and Tobias Feger ({\it right}). The final device will be mounted in a solid casing. Note the wooden rack ({\it left}) for easy and fast geometrical improvements.}
\end{figure}

\begin{figure}[ht]
\centering
\includegraphics[height=5cm]{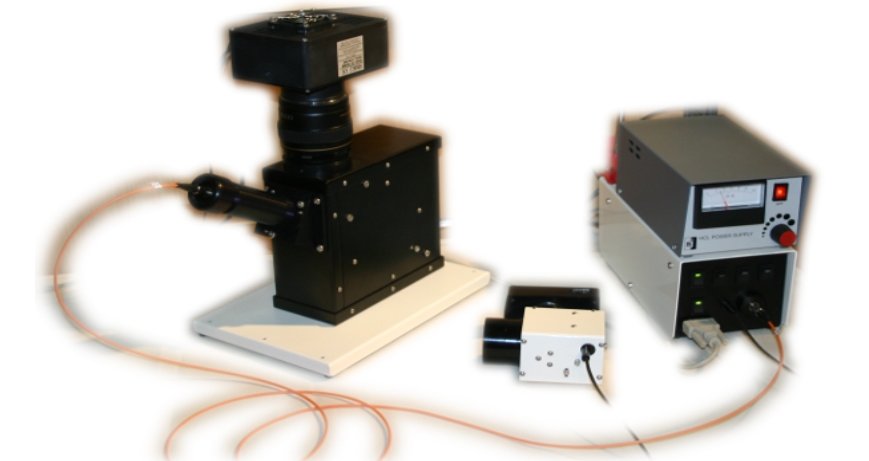}
\caption{\label{eShel}
Off-the-shelf fiber-fed Echelle with prism cross-disperser from Shelyak.}
\end{figure}

In the past few years, various spectrographs have been successfully designed and constructed by
dedicated amateur astronomers using off-the-shelf optics and blazed gratings, and have been properly
adapted to respective telescopes. Two examples using standard photographic optics are shown in Fig.~\ref{specs}. 
The first generation of instruments delivering a spectral resolution
of more than 10.000 are now also available commercially. The most popular device is the LHIRES III Littrow spectrograph by the company Shelyak in France (Fig.~\ref{lhires}). Because of its simple plug-and-play design for typical amateur telescopes like Celestron and Meade it is widely used for different campaigns (see below). Above this, skilled amateurs now start to design their own Echelle spectrographs. Two examples are shown in Fig.~\ref{stoberfeger}. 
Today, the first off-the-shelf Echelle spectrographs are also
available, including a complete and tested software routine for a ``plug-and-play" data reduction chain 
(Fig.~\ref{eShel})\footnote{www.shelyak.com}. 
Amateur and off-the-shelf Echelle spectrographs have the same performance but are often cheaper than similar professional prototypes for small telescopes. This is mainly due to commercial off-the-shelf serial production. 


\section{Long-term campaigns, surveys, monitoring}

Amateur spectroscopic equipment can easily be used for scientific investigations of stellar physics,
particularly the study of bright emission line stars where line profile analysis of their often fast
varying spectra can be performed. For instance, using a standard 10\,inch telescope, a signal-to-noise ratio (S/N) of about
100 can be achieved within 30\,minutes for a star of about 8 magnitudes in $V$ and for a two pixel resolution
of about 1\,{\AA}. Objects of the order of $V=$10\,mag and fainter are generally excluded due to limited amateur telescope
apertures, although with longer exposure times and/or lower S/N this limit can be extended to even fainter stars. 
Hence, amateur spectroscopists can fill specific gaps for detailed investigations.
These are \textbf{A)} \textbf{spectroscopic long-term campaigns} monitoring line profiles for periods of the order of months
or even years, \textbf{B) surveys} to support detailed observations by large or space-based telescopes and
\textbf{C) monitoring} of specific spectroscopic parameters over many years.

\subsection{The long-term $\epsilon$ Aurigae campaign}

\begin{figure}[ht]
\centering
\includegraphics[height=5cm]{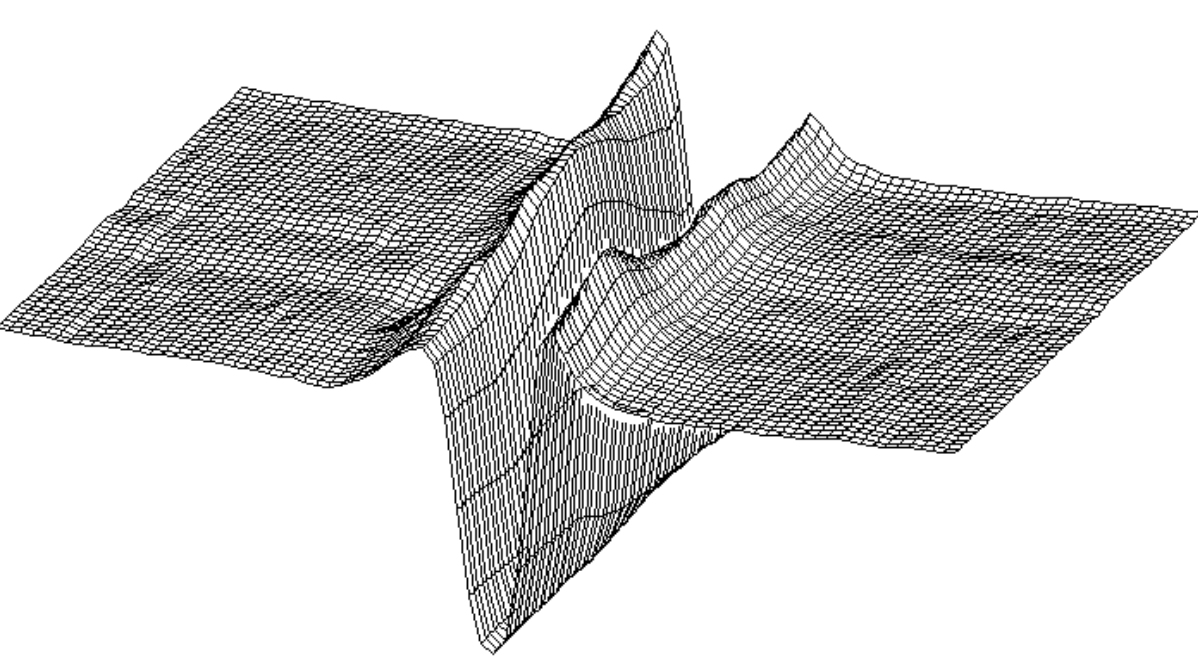}
\caption{\label{buil2}
H$\alpha$ time series for the eclipsing binary $\epsilon$ Aur between July 21, 2009 and March 10, 2010 with 3-days increment. Celestron C11, Shelyak Echelle spectrograph, average spectral resolving power R = 11000.}
\end{figure}

A prominent example of a long-term amateur campaign is the eclipsing binary $\epsilon$ Aurigae (F0Ia +
companion) with an orbital period of about 27 years\footnote{http://www.threehillsobservatory.co.uk/epsaur\_spectra.htm}. 
The respective campaign group\footnote{http://www.hposoft.com/Campaign09.html} is internationally acting for the 2009 - 2011 eclipse of the star system. A periodic campaign newsletter regularly highlights recent events.
H$\alpha$ time series, obtained by Christian Buil from his balcony in Marseille using a Shelyak Echelle spectrograph (Fig.~\ref{eShel}) is shown in Fig.~\ref{buil2}. Because of its long period of 27 years $\epsilon$ Aurigae is simply a prototype object for amateur observations, impossible to perform by pro's on such a timescale (see also R.~Leadbeater, these proceedings). 

\subsection{The long-term Mons campaign}
\label{wr140}

\begin{figure}[ht]
\centering
\includegraphics[height=5cm]{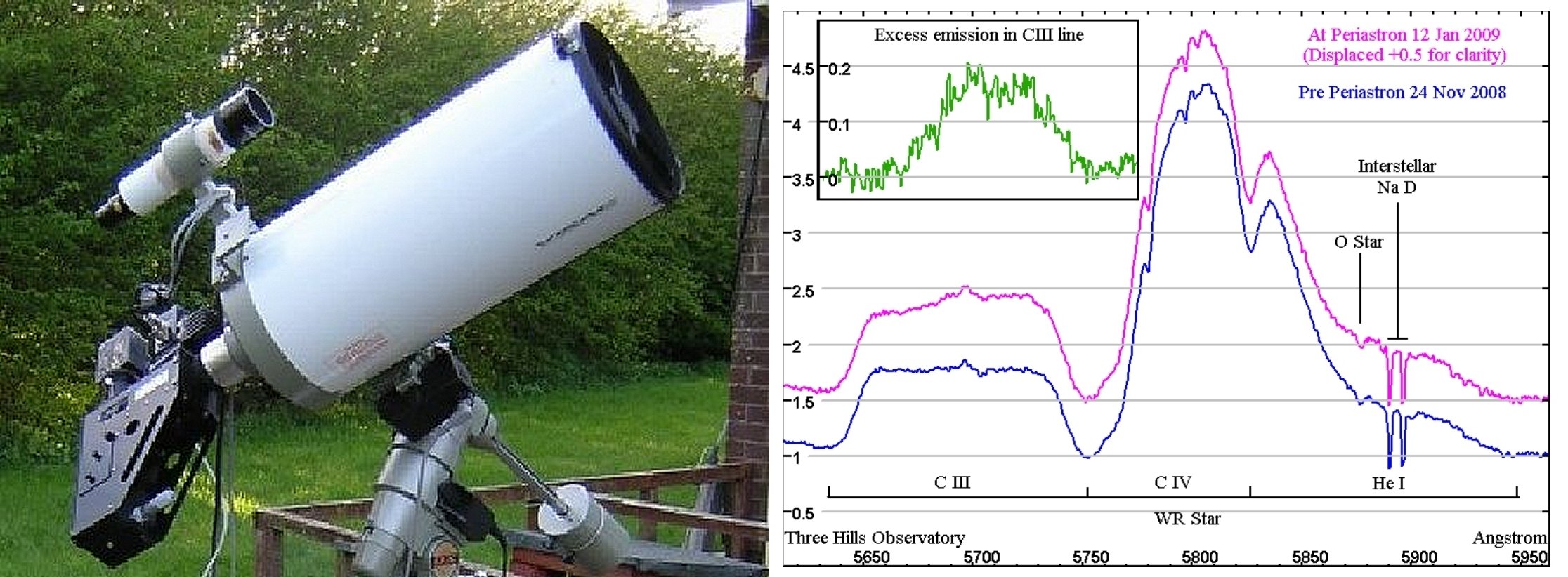}
\caption{\label{lead}
{\it Left}: The private Three Hills Observatory of Robin Leadbeater in Cumbria, England, consisting of a Vixen 20\,cm Cassegrain and an off-the-shelf LHIRES III Littrow slit spectrograph. {\it Right}: Spectral variability of WR~140 within seven weeks obtained with the left instrument. }
\end{figure}

The archetype of colliding-wind binary systems is the 7.9-year period and highly excentric WR+O binary system WR~140 (HD193793). Twenty-six amateurs and professionals from eight countries observed the prominent C{\sc iii} wind line and its excess during periastron passage from Tenerife and from various home observatory in Europe to estimate the ephemeris of the system. All stations used the LHIRES III spectrograph. As part of this global campaign, Robin Leadbeater obtained spectra during periastron passage from his home observatory (Fig.~\ref{lead}). Figure~\ref{lead} also shows his two spectra of C{\sc iii}/C{\sc iv} 
before and during periastron passage obtained with his configuration. The inset shows the resulting excess emission due to the wind-wind interaction shock cone. 

\subsection{Surveys}

\begin{figure}[ht]
\centering
\includegraphics[height=5cm]{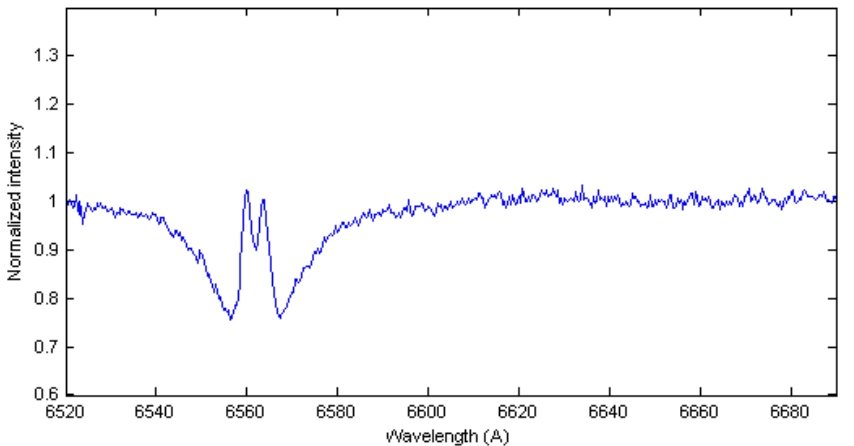}
\caption{\label{corot}
H$\alpha$ amateur {\it COROT} survey spectrum of the $V=$6.14 B9Ve star HD194244, obtained with a Celestron C11 and an LHIRES III spectrograph with 2700\,s exposure time.  }
\end{figure}

\begin{figure}[ht]
\centering
\includegraphics[height=5cm]{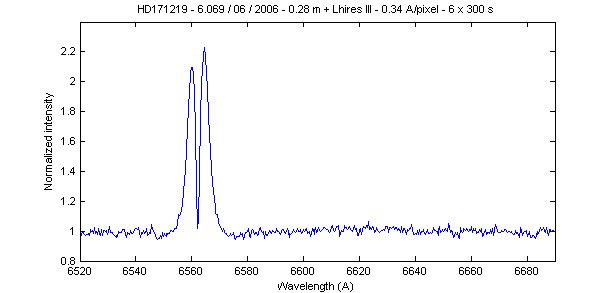}
\caption{\label{171219}
H$\alpha$ amateur {\it COROT} survey spectrum of the $V=$7.65 B5IIIe star HD171219. Instrumental setup as above, but 1800\,s exposure time.  }
\end{figure}

A "classical" example for an astronomical survey, supported by amateur observers, has been the astrometric {\it High Precision Parallax Collecting Satellite} ({\it Hipparcos}), launched in 1989. Amateur astrometry has been performed for centuries and dedicated amateurs had already the respective experience to obtain high precision data. As a result, many observers contributed their measurements to perform a successful satellite project.  Such ProAm surveys are today possible in spectroscopy, as well. 
The presently most popular spectroscopic survey is the {\it COROT} Be Stars Survey\footnote{http://www.astrosurf.org/buil/corot/data.htm} project for the astroseismology satellite {\it COROT} ({\it COnvection, ROtation and planetary Transits}). A respective amateur {\it COROT} survey of bright stars (e.g., Be stars), as performed under professional supervision can help understanding spectral variability like non-radial pulsations or oscillations in the respective Be star disks. Example spectrA, obtained by a French amateur group around C. Buil, are shown in Figs.~\ref{corot} and~\ref{171219}.

\subsection{Monitoring}

\begin{figure}[ht]
\centering
\includegraphics[height=7cm]{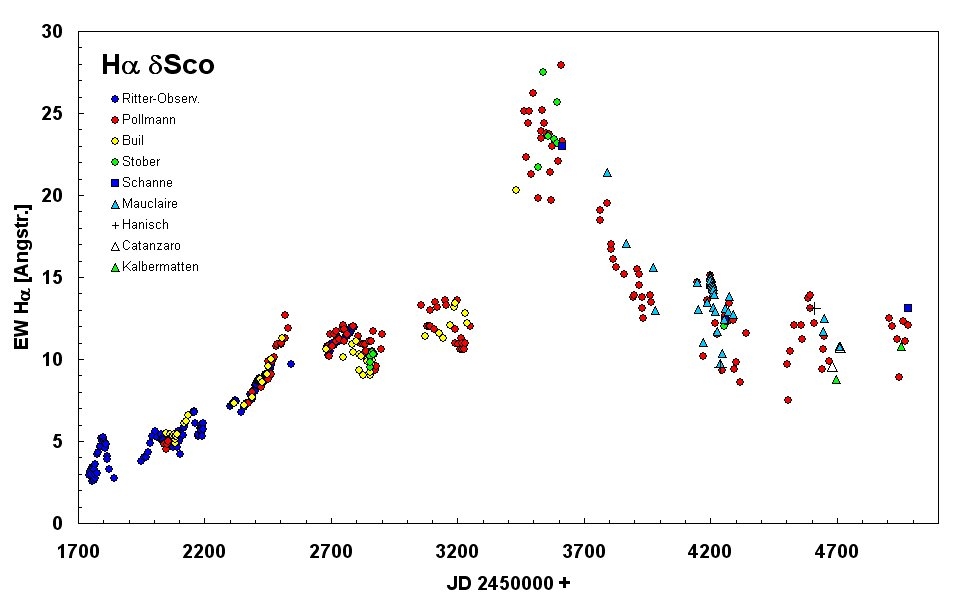}
\caption{\label{delsco1}
H$\alpha$ equivalent width measurements for the Be star $\delta$ Scorpii over about nine years. Note that the first measurements taken by professionals (blue circles) have been dramatically extended by a group of amateur astronomers.  }
\end{figure}

The most problematic task for professional spectroscopy is probably extremely long-term monitoring
of specific spectral parameters. Only snapshots within relatively short 
time-scales are usually possible, resulting in large time gaps. The true long-term description of the
physical behaviour remains hidden. Delays in publishing results, sometimes for several years, do not
match the regular needs of a pro. Continuous monitoring is not a priority in professional
spectroscopy. This, however, is one of the cornerstones of amateur work using simple standard
procedures (e.g., equivalent width and radial velocity measurements) combined with well-known
equipment and good routines. An example is shown in Fig.~\ref{delsco1} for the Be star $\delta$ Scorpii. 
This Be star binary with an orbital period of about 11 years is one of the key targets of amateur work and for numerous observations. The next periastron passage will take place in July 2011 and is already in the focus of an extended ProAm campaign, including observations from Teide observatory, as for WR~140 (see Section \ref{wr140}).

\section{How to establish a ProAm campaign}

If the professional community wants to take advantage of amateur measurements one should keep some basic issues in focus. Instrumental knowledge and observational hands-on skills are already present in the amateur domain. On the other hand,  scientific knowledge (e.g., physics, procedures, data interpretation) need to be contributed by those who have a complete university education in this field and already have the relevant experience. The professional community can not expect
complete campaign proposals from amateur astronomers but should first take their specific spectroscopic
needs to the amateurs and discuss them. For instance, the Mons campaign on WR~140 took only place because of a close contact between a professional scientist (Tony Moffat) and one of his previous students who is now active in the German amateur community. The {\it COROT} Be stars survey project is mainly driven by a professional group\footnote{http://www.ster.kuleuven.be/~coralie/members.htm} working closely together with amateurs in France\footnote{http://astrosurf.com/aras}, Germany\footnote{http://spektroskopie.fg-vds.de} and elsewhere.  
Potential ProAm campaigns need some basic details. After an announcement in the respective communities it is essential to give information about the physical background and basic parameters (e.g., S/N, spectral resolution, etc.) to all interested observers. This is best done by a respective website. Unfortunately, there are only two well established communities of significant size, namely in France and Germany. In these two communities, respective discussion forums are available and it is recommended to use them for proper discussion. For the campaign management it is also recommended to separate science from administration issues. A highlight of each campaign is potential observing time at a professional observatory. Amateurs regularly do not have such access but are generally highly interested to go for it---often even at their own expense (if limited), as happened for the Mons campaign. Finally one should find some minimum financial resources (depending on the observatory site) to cover potential financial deficits (higher equipment transportation costs, unforeseen events, etc.).

\section{Future plans and Summary} 

In May 2010 most of the key players in the WR~140 MONS campaign organised a wrap-up meeting at Convento da Arr\'abida\footnote{http://astrosurf.com/joseribeiro/e\_arrabida.htm} close to Lisbon, of which these are the proceedings. The group, now called ``ConVento", will establish an informal website covering future ProAm campaigns, respective information about specific stellar targets and a mailing list / discussion forum. Every interested spectroscopist and photometrist is invited to join the group. The link to this website will soon be found at www.stsci.de.
Considering the present situation in astronomy it seems obvious that skilled and sophisticated amateurs equipped with state-of-the art instrumentation in their domain, can successfully contribute their knowledge and enthusiasm to modern spectroscopic campaigns, either at their home observatories or at professional sites. The only obstacle to making continuous observations like those at professional sites is
the local weather and the fact that amateur astronomer usually have to work in their daily job. This
however can be circumnavigated by joint campaigns, as shown above. It is up to the professional
community to uncover this valuable treasure.

%
%
\section*{Acknowledgements}
I thank Tony Moffat, Thierry Morel and Gregor Rauw for their lasting support. 

\newpage
\qquad

\end{document}